\documentclass[aps, prd, reprint, groupaddress]{revtex4-1}
\usepackage[pdftex]{graphicx}% Include figure files
\usepackage{dcolumn}% Align table columns on decimal point
\usepackage{bm}% bold math
\usepackage{latexsym}
\usepackage{amsmath}

\usepackage{amssymb, amsthm}
\usepackage{mathrsfs}

\begin{document}

\title{Ground Based Low-Frequency Gravitational-wave Detector With Multiple Outputs}%

\author{Ayaka Shoda$^1$}
\email{ayaka.shoda@nao.ac.jp}
\author{Yuya Kuwahara$^2$}
\author{Masaki Ando$^{1,2,3}$}
\author{Kazunari Eda$^{2,3}$}
\author{Kodai Tejima$^2$}
\author{Yoichi Aso$^1$}
\author{Yousuke Itoh$^3$}%
\affiliation{$^1$ Gravitational Wave Project Office, Optical and Infrared Astronomy Division,
National Astronomical Observatory, Osawa 2-21-1, Mitaka, Tokyo 181-8588, Japan \\
$^2$ Department of Physics, Graduate School of Science, University of Tokyo, Hongo 3-7-1, Tokyo 113-0033, Japan \\
$^3$ Research Center for the Early Universe (RESCEU), Graduate School of Science, University of Tokyo, Hongo 3-7-1, Tokyo 113-0033, Japan}

%\affiliation{$^2$ Department of Physics, Graduate School of Science, Kyoto
% University, Kyoto 606-8502, Japan}
%\affiliation{$^2$ Research Center for Neutrino Science, Tohoku University, Sendai 980-8578, Japan} %

\begin{abstract}
We have developed a new gravitaional-wave (GW) detector, TOrsion-Bar Antenna (TOBA), with multiple-output configuration.
TOBA is a detector with bar-shaped test masses that rotate by the tidal force of the GWs.
In our detector, three independent information about the GW signals can be derived by monitoring multiple rotational degrees of freedom, i.e., horizontal rotations and vertical rotations of the bars.
Since the three outputs have different antenna pattern functions, the multi-output system improves the detection rate and the parameter estimation accuracy.
It is effective in order to obtain further details of the GW sources, such as population and directions. 
We successfully operated the multi-output detector continuously for more than 24 hours with stable data quality.
Also, the sensitivity of one of the signals is improved to be $1 \times 10^{-10}$ ${\rm Hz}^{-1/2}$ at 3 Hz by the combination of the passive and active vibration isolation systems, while sensitivities to possible GW signals derived from the vertical rotations are worse than that from the horizontal rotation.
\end{abstract}

\maketitle

\section{Introduction}
On September 2015, the gravitational-wave (GW) was directly detected by the Advanced Laser Interferometer Gravitational-wave Observatory (aLIGO) for the first time \cite{Abbott2016}.
The detected GW, labeled by GW150914, was the signal from the inspiral and merger of two black holes with the masses of about 30 $M_{\odot}$.
The signal was the first evidence of the binary black hole merger.
As another event of black hole merger, GW151226, had been detected \cite{Abbott2016_151226}, further details of binary black hole systems, such as distribution and population of black holes, might be revealed by further observations.

While aLIGO detected the transients with the durations of a few hundreds milliseconds, a longer observations of the binary systems would provide much information of the binary systems, such as spins of black holes.
Binary systems with the masses of about 30 $M_{\odot}$ emit GWs with the frequencies of about 0.1 Hz at 15 days before the merger.
Such a long term observation of the GW signals from the binary systems would effectively solve the degeneracy of the parameters, such as their spins, which is expected to be a clue about the evolution of the black holes \cite{Abbott2016ApJ, Nakamura2016}.
In addition to the black holes with the masses of the order of 10 $M_{\odot}$, it is important to search for black holes with various masses.
Observation of various black hole mergers would elucidate the evolution process of super-massive black holes.
In order to observe the GW signals from binary systems with various masses, it is necessary to observe GWs at various frequencies. 
Binaries with heavier masses emit GWs at lower frequencies than the observation band of the interferometers. 

A TOrsion-Bar Antenna (TOBA) \cite{Ando2010} is a gravitaional-wave (GW) detector that is sensitive to GWs at around 1 Hz, while the observation band of interferometric GW detectors, such as LIGO, are above about 10 Hz.
One of the astronomical targets of TOBA is intermediate-mass black hole binaries.
The space-borne interferometric detectors, such as Laser Interferometer Space Antenna (LISA) \cite{Vitale2014} and DECI-herz Gravitational-wave Observatory (DECIGO) \cite{Kawamura2011}, would also have the observation frequency band below 10 Hz.
Comparing to such detectors, TOBA can be realized on the ground with an accessibility for repair and upgrade, and without risk for large cost.
Though the final target sensitivity of TOBA is not as good as space detectors, the observation range is expected to be as far as 10 Gpc with the luminosity distance for the intermediate-mass black hole mergers.
TOBA has two bar-shaped test masses which are suspended at their centers. 
GWs are detected by monitoring their relative rotation excited by the tidal force from GWs.
Since the resonant frequencies of the test masses in the torsional modes are as low as a few mHz, TOBA has sensitivity at low frequencies even on the ground.
As a proof of concept, the first prototype of TOBA had been developed with a single test mass bar \cite{Ishidoshiro2010} and set the first upper limit on a stochastic GW background from 0.03 to 0.8 Hz \cite{Ishidoshiro2011, Shoda2014}.

On the other hand, it is necessary to use more than three detectors in order to determine the parameters of binaries, such as masses of the objects, polarization angle, and the sky position of the source because the conventional detectors have poor directivity \cite{Schutz2011}.
Therefore, we proposed {\it the Multi-output TOBA}, which provides three independent signals from a single detector by monitoring multiple rotational degrees of freedom of the test masses \cite{Eda2014}.
The multi-output system improves the event rate and the angular resolution, which would enhance the low-frequency GW astronomy even with fewer detectors.

In this paper, we introduce the first Multi-output TOBA detector.
Its main feature is the new suspension system for the multi-output configuration, which also performs as a vibration isolation system.
In the section \ref{sec:TOBA}, we explain its principle and the target sensitivity.
The detector configuration that we developed is described in the section \ref{sec:Detector}.
Its characteristics is mentioned in the following section.

\section{TOBA}\label{sec:TOBA}
\subsection{Principle of GW detection using TOBA}
\begin{figure}[tb]
\centering
\includegraphics[width=70mm]{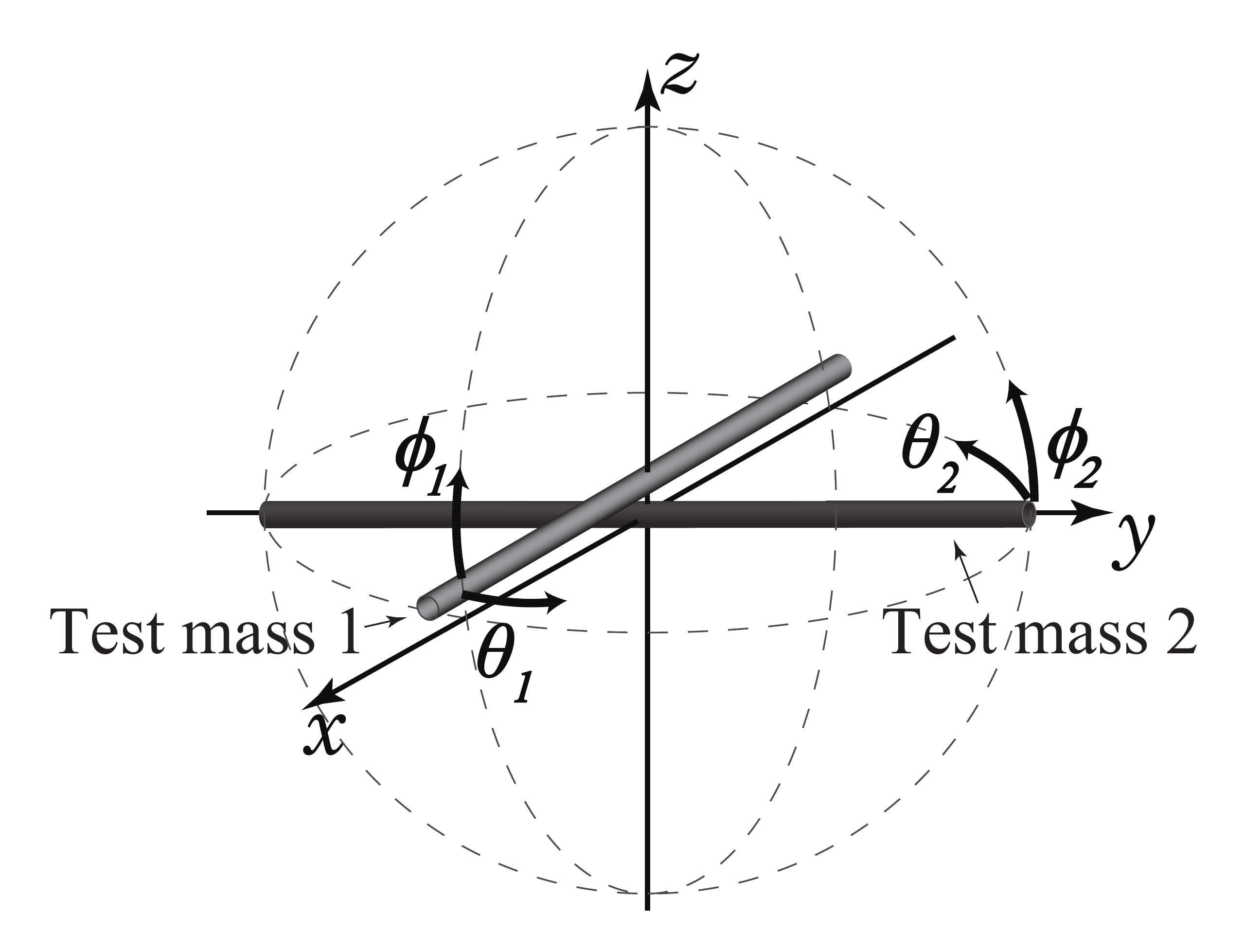}
\caption{\label{fig:TOBA}Schematic view of the TOBA test masses.}
\end{figure}

TOBA has two bar-shaped test masses that rotate differentially by the tidal force from GWs as shown in Fig. \ref{fig:TOBA} \cite{Ando2010}.
The equation of the motion of the test mass bar in the rotational mode is 
\begin{equation}\label{eq:eom}
I_z \ddot{\theta}(t) + \gamma_z \dot{\theta}(t) + \kappa_z \theta (t) = \frac{1}{4}\ddot{h}_{ij}q_z^{ij},
\end{equation}
where $\theta, I_z , \gamma_z,$ and $\kappa_z$ are angular fluctuation, the moment of inertia, the damping constant, and the spring constant around $z$-axis, respectively.
$h_{ij}$ is the amplitude of the GW, and $q_z^{ij}$ is the dynamical quadrupole moment of the test mass for the rotation along the $z$ axis.
The frequency response of the angular fluctuation of the test mass is derived from Eq. (\ref{eq:eom}) as
\begin{eqnarray}
\tilde{\theta}(\omega) &=& \sum_{A=+,\times} H_A(\omega)\tilde{h}_A , \\
H_{A}(\omega) &=& \frac{q_A}{2I}\frac{\omega^2}{\omega^2 - \omega_0^2(1+i\varphi) + i\omega\gamma_{z}/I}\label{eq:H},
\end{eqnarray}
where $\varphi$ and $\tilde{h}_A$ are the loss angle and the amplitude of the GW coming along $z$-axis with the polarizations of $A=+$ and $\times$, respectively.
$\omega_0 = \sqrt{\kappa_z/I}$ is the resonant frequency of the torsional mode, above which the angular fluctuation due to the GW is approximated to be independent from its frequency.

%\begin{figure}[tb]
%\centering
%\includegraphics[width=70mm]{Ha.pdf}
%\caption{\label{fig:GWTF}Transfer function from the GW amplitude to the angular fluctuation of the test mass. Here, the resonant frequency $f_0 = \sqrt{\kappa_z/I_z}/2\pi$ is assumed to be 0.1 Hz.}
%\end{figure}

In the multi-output system, we also consider rotations of the bars on the vertical planes.
Considering the test mass 1 in Fig. \ref{fig:TOBA}, the angular fluctuation along the $y$-axis also obeys the similar equation as Eq. (\ref{eq:eom}):
\begin{equation}I_y \ddot{\phi_1}(t) + \gamma_y \dot{\phi_1}(t) + \kappa_y \phi_1 (t) = \frac{1}{4}\ddot{h}_{ij}q_y^{ij}.
\end{equation}
It means that the bar also rotates vertically due to GWs coming along $y$-axis.
Therefore, we can derive two independent signals $\theta$ and $\phi$ from the single test mass bar.

Since we have two orthogonal test mass bars, it is possible to derive three independent signals from the single detector, i.e., $\theta = \theta_1 = -\theta_2$, $\phi_1$, and $\phi_2$, where the suffix indicates the two test masses.
The sensitivity to GW signal derived from $\theta$, calculated as $\theta = (\theta_1 - \theta_2)/2$, would be better by $\sqrt{2}$ than the sensitivity derived only from one of the rotational signals, when the noise appeared in $\theta_1, \theta_2, \phi_1$, and $\phi_2$ are un-correlated and the same level.
This multi-output system would increase the expected detection rate by about 1.7 times, since the three signals have different sensitive areas in the sky. 
Also, parameter estimation accuracy for short-duration signals would be improved since the three independent information helps to break the degeneracies of the parameters \cite{Eda2014}.

\subsection{Target Sensitivity and the Previous Research}
The final target sensitivity of TOBA is about $1 \times 10^{-19}$ ${\rm Hz}^{-1/2}$ in strain at 1 Hz as described in \cite{Ando2010}, which is limited mainly by the shot noise, the radiation pressure noise, and gravity gradient noise.
This sensitivity can be achieved by sensing the rotation of large cryogenic test mass bars with the length of 10 m, using Fabry-Perot interferometer.
The test mass bars and the suspension wires should be cooled down in order to reduce the thermal noise.
Since several advanced technologies are necessary for the final TOBA configuration, it is required to develop each component using prototypes.

The first prototype TOBA had been developed previously \cite{Ishidoshiro2010}.
The first prototype has a single bar shaped test mass with a magnetic levitation system in order to suspend the test mass softly in the rotational degree of freedom.
The test mass has the length of 20 cm, and its horizontal rotation is measured by a Michelson interferometer, which means that it has the single-output configuration.
It successfully tested the basic principle of TOBA, and set the first upper limit on a stochastic gravitational-wave background at 0.2 Hz \cite{Ishidoshiro2011}.
The sensitivity of the first prototype TOBA is limited by the magnetic noise induced by the magnetic suspension system, and the seismic noise coupling.

As a next step, we have developed a Multi-output TOBA, as described in the following sections.
Its main target is the development of the suspension system that realizes the multi-output system.
Also, the passive and active vibration isolation systems attenuate the noise caused by the seismic motion.

\section{Multi-output TOBA Detector}\label{sec:Detector}
\subsection{Overview}
\begin{figure}[tb]
\centering
\includegraphics[width=70mm]{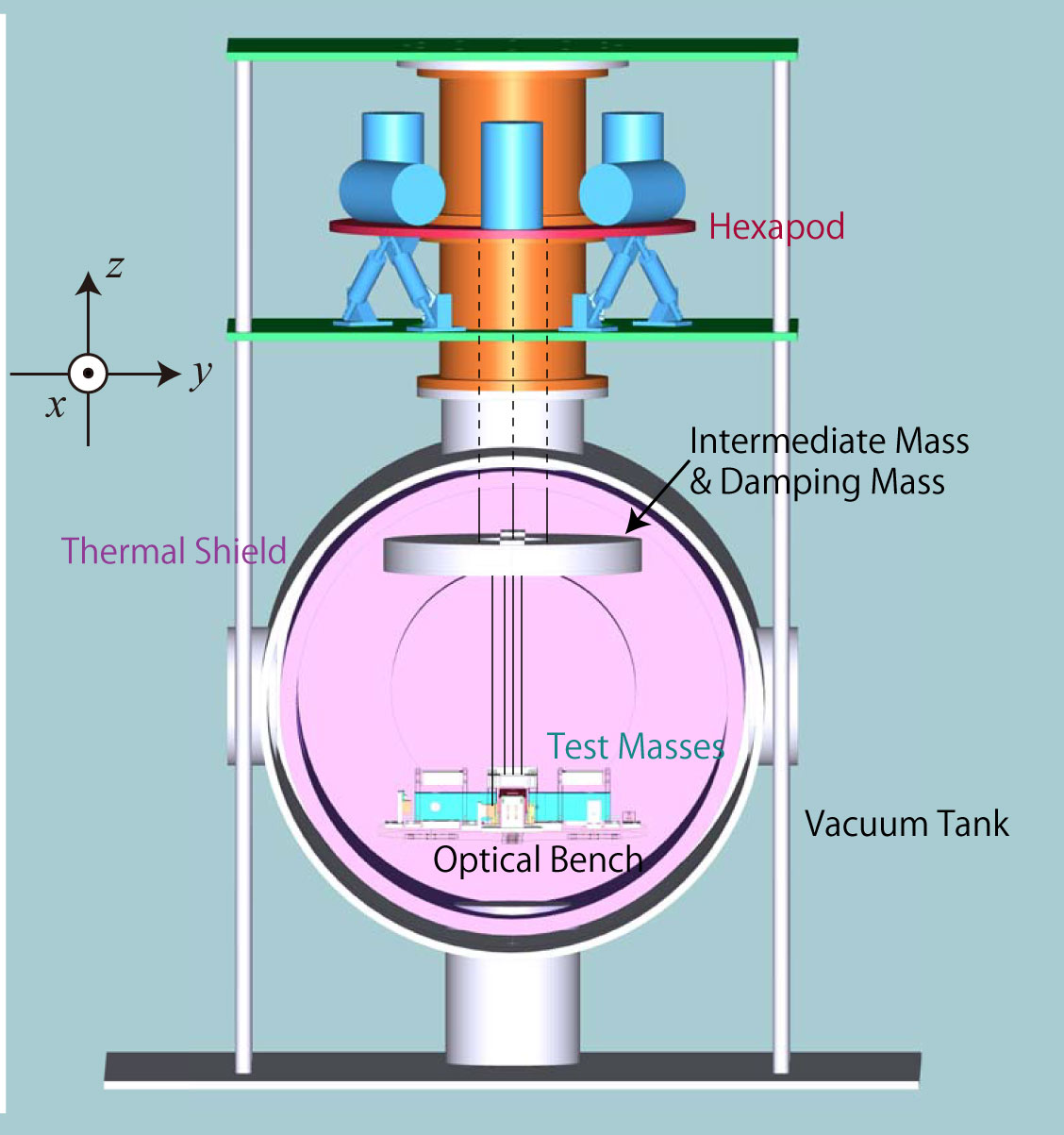}
\caption{\label{fig:schemview} (color online). Overview of the multi-output TOBA.}
\end{figure}
The schematic view of the multi-output TOBA is shown in Fig. \ref{fig:schemview}. 
The two orthogonal test masses that sense the GWs are suspended from a Hexapod-type active vibration isolation table (AVIT) via an intermediate mass, which is magnetically damped by a damping mass. 
The optical bench where the sensors and actuators are set is also suspended from the intermediate mass.
The suspension system except the actuators and sensors of the AVIT is in a vacuum tank.

\subsection{Test Masses}\label{sec:TM}
\begin{figure}[tb]
\centering
\includegraphics[width=70mm]{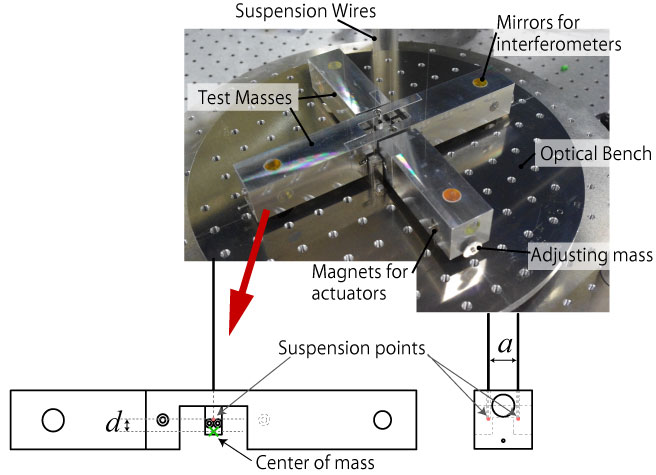}
\caption{\label{fig:testmass} (color online). Picture of the test masses. Each test mass is suspended by two wires so that the center of each mass can be located at the same position.}
\end{figure}

The picture of the test masses is shown in Fig. \ref{fig:testmass}. The test masses are designed for the multi-output system and the test of common mode noise reduction. 
The two orthogonal test mass bars with the length of 24 cm are suspended by the two parallel wires respectively, so that the centers of the masses can be located at the same position in order to maximize the common mode noise reduction rate in horizontal rotation signal.
This design is implemented since it may reduce the noise caused from the common rotational displacements of the bars.
The common mode noise reduction rate is expected to be large when the sensitivity is limited by the environmental disturbance that effects the test mass rotations commonly, while the reduction rate is $\sqrt{2}$ in strain at the minimum when the noise sources of the two signals are independent.

The resonant frequency in the horizontal rotational mode is $f_{\theta 0} = \sqrt{mga^2/Il}/2\pi$, where $m, g$, $a$, $I$ and $l$ are the mass of the test mass, the acceleration of gravity, the distance between the two suspension wires as shown in Fig.\ref{fig:testmass}, moment of inertia of the test mass, and the length of the wire, respectively.
The resonant frequency in the vertical rotational mode is written as $f_{\phi 0} =\sqrt{mgd/I}/2\pi$, where $d$ is the height distance between the suspension points and the center of mass as shown in Fig.\ref{fig:testmass}. 
The suspension points of the test masses are set to be close to their center of mass in order to minimize the resonant frequency of the vertical rotation. 
In our set up, the resonant frequency of the horizontal and vertical rotational modes are 0.10 Hz and 0.15 Hz, respectively.
The resonant frequencies are set at around 100 mHz in our prototype in order to realize the new suspension system with the compact setup.
The resonant frequencies would be pushed down by using larger test masses in the future upgrade in order to widen the observation band.

\subsection{Readout System}
\begin{figure}[tb]
\centering
\includegraphics[width=70mm]{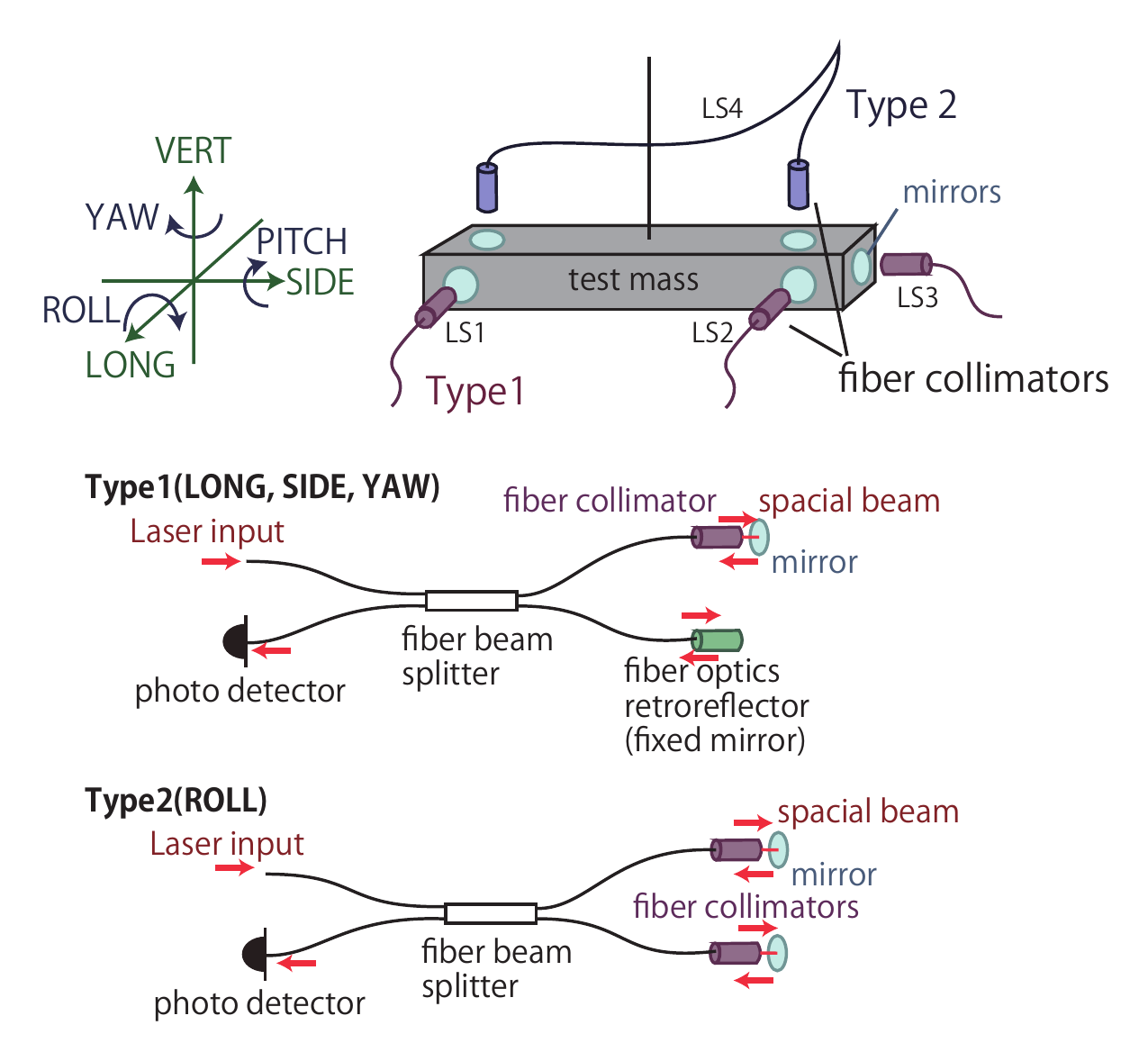}
\caption{\label{fig:sensor}(color online). Configuration of the fiber laser interferometers.}
\end{figure}

The main sensors that we used for the observation in the Multi-output TOBA are the Michelson interferometeric sensors. 
The motion of the bar is monitored by measuring the phase shift of the beam reflected by the mirrors attached at the bar.
Since it is necessary to have several sensors around the test masses in order to monitor three independent rotational signals, the fiber optics are used for the space saving.
The sensor configuration is shown in Fig. \ref{fig:sensor}.
The type-1 interferometers, which measure the displacement of each mirror attached on the test masses, monitor the yaw, longitudinal, and side motion of the bar.
The type-2 interferometers that sense the differential displacement of the two end mirrors are set as sensors for the roll motion.
The position of the test masses are controlled by the coil-magnet actuators so that the fringe of the interferometer can be kept at their middle.
The GW signals are derived from the feedback signal on the actuators.
The test masses are controlled in longitudinal, side, yaw, and roll modes.
These sensors are set on the suspended optical table explained in the next sub-section.

\subsection{Vibration Isolation System}
Though the seismic motion in the rotational degree of freedom is small, the seismic vibration isolation system is necessary since the translational vibration couples to the rotational signals.
For example, non-parallel mirrors at the both ends of the test mass induce the translational seismic noise coupling \cite{IshidoshiroThesis, ShodaThesis}. 
The test masses, and the optical table where the sensors and actuators are set, are suspended in order to attenuate the seismic motion above resonant frequencies of those pendulum modes.
However, since the resonant frequencies of the pendulum modes are about 1 Hz, an additional seismic isolation system for low frequencies are necessary for TOBA.
Therefore, we developed the active vibration isolation table (AVIT) for low-frequency vibration isolation and the whole system is suspended from the AVIT.

\begin{figure}[tb]
\centering
\includegraphics[width=70mm]{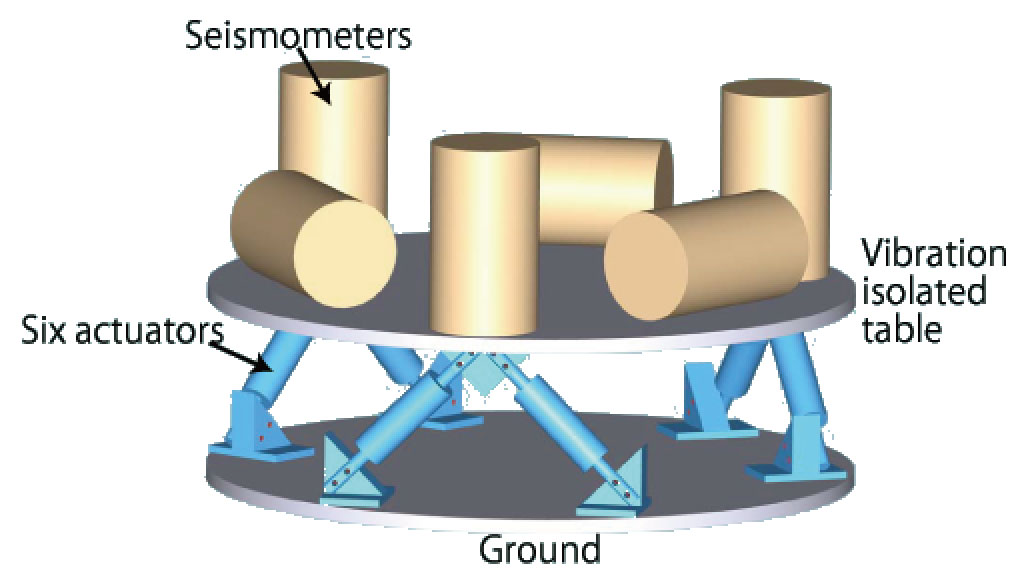}
\caption{\label{fig:Hexa}(color online). Schematic view of the Hexapod-type active vibration isolation table (AVIT). The whole suspension system is suspended from the top table. It has six piezoelastric actuators as legs so that they can actuate the top table in six degrees of freedom. The six seismometers are set in order to sense the vibration of the top table.}
\end{figure}
The AVIT is a table with six legs composed of piezoelastic elements (PZTs) as shown in Fig. \ref{fig:Hexa}.
The PZTs (P844.30, products of Phisik Instrumente) move the position of the top plate. 
Those PZTs have tips with slits at the both ends in order to avoid non-linear effect when they push or pull the table.
The vibration of the top table is suppressed by the feedback control using six seismometers (L-4C geophones, products of Sercel) set on the top plate and the PZTs.
Note that reflective type position sensors are used in order to measure the DC position of the top table relative to the ground since the lack of the sensitivity of seismometers at low frequencies causes drift of the top table.

\begin{figure}[tb]
\centering
\includegraphics[width=92mm]{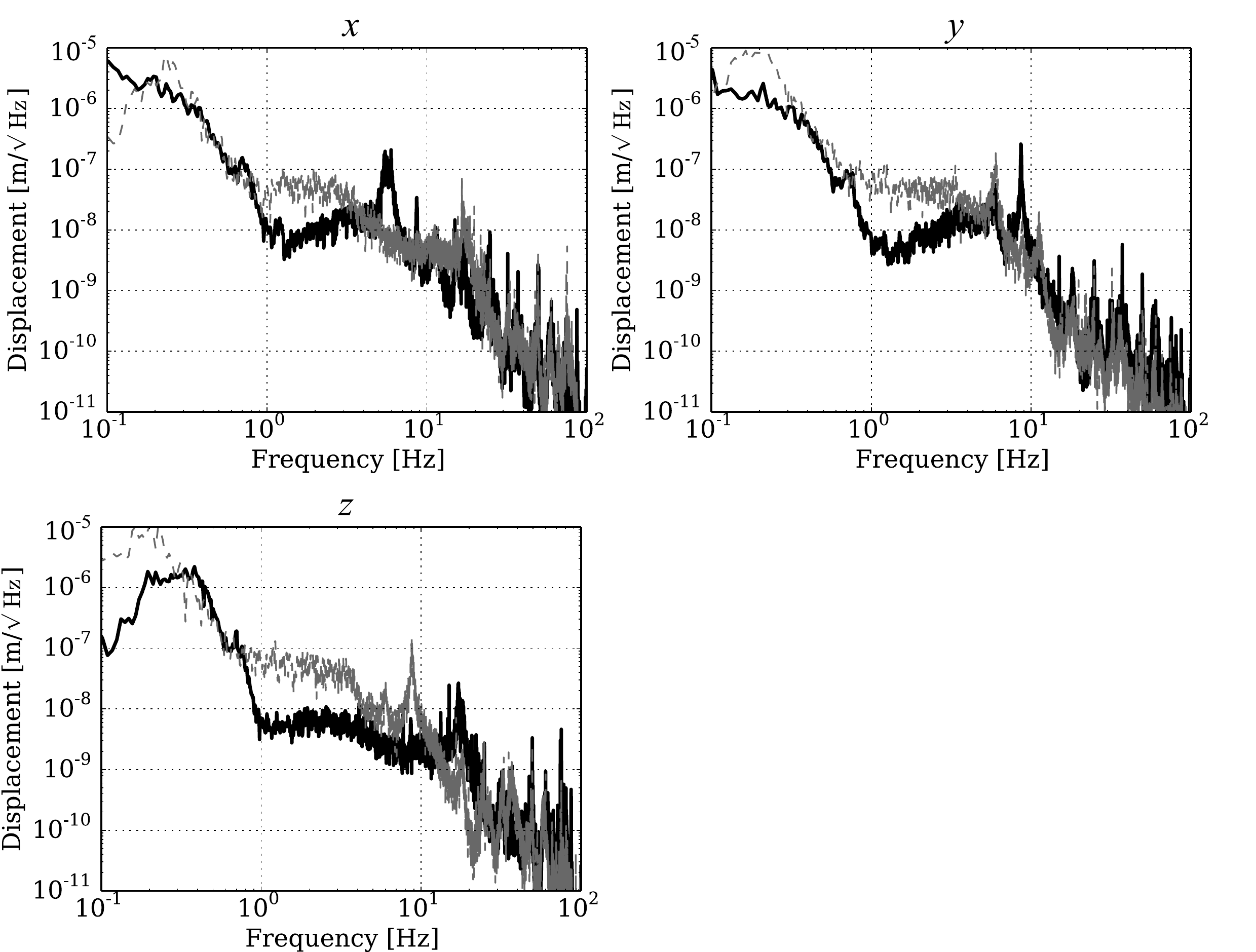}
\caption{\label{fig:seis_hexa} Seismic isolation performance of the AVIT in each degrees of freedom. The dashed lines represent the seismic displacement with the AVIT off, and the solid lines represent one with the AVIT on. The coordinate is defined in Fig. \ref{fig:schemview}.}
\end{figure}

The seismic displacement of the top table is shown in Fig. \ref{fig:seis_hexa}.
The vibration isolation ratio from the ground motion at 1 Hz is achieved to be almost 10 times.
Its performance is mainly limited by the range of the PZTs and the resonance of the frame where the AVIT is sitting.

The AVIT can attenuate the vibration at around 1 Hz even with its compact body of 45 cm in diameter, while the passive vibration isolation requires large setup for low-frequency vibration isolation, such as inverted pendulums \cite{Takamori2007, Matichard2015, Braccini1993}.
Also the AVIT is effective to attenuate the vibration of the heat link in the cryogenic system for thermal noise reduction that is planned to be implemented in the future upgrade.
It is because the rigid structure enables the AVIT to suppress the vibration induced directly into the vibration attenuated plate, i.e., the vibration of the heat link attached at the suspension point.

\section{Detector Characterization}\label{sec:Characteristics}
\subsection{Calibration}
The angular fluctuations of the test masses are read by the laser interferometers.
The angles are calibrated into the GW signal outputs as follows:
\begin{eqnarray}
s_1 &=& \frac{1}{H_{\times 1}}\frac{1}{2} (\theta_1 - \theta_2), \\
s_2 &=& \frac{1}{H_{\times 2}}\phi_1, \\
s_3 &=& \frac{1}{H_{\times 3}}\phi_2,
\end{eqnarray}
where $H_{\times i}$ $(i = 1,2,3)$ are defined as
\begin{equation}
H_{\times i} = H_{\times i}^{\rm TM} - H_{\times i}^{\rm OB}.
\end{equation}
$H_{\times i}^{\rm TM}$ and $H_{\times i}^{\rm OB}$ are the transfer functions from the GW signal to the angular fluctuations of the test masses and the optical bench derived from Eq. (\ref{eq:H}), respectively.
Here, $H_{+i}$ are considered to be zeros since $q_{+}=0$ when the test masses are suspended along $x$-axis and $y$-axis.
Note that the optical bench is also sensitive to GWs since it is also suspended from the intermediate mass.

\begin{figure}[tb]
\centering
\includegraphics[width=70mm]{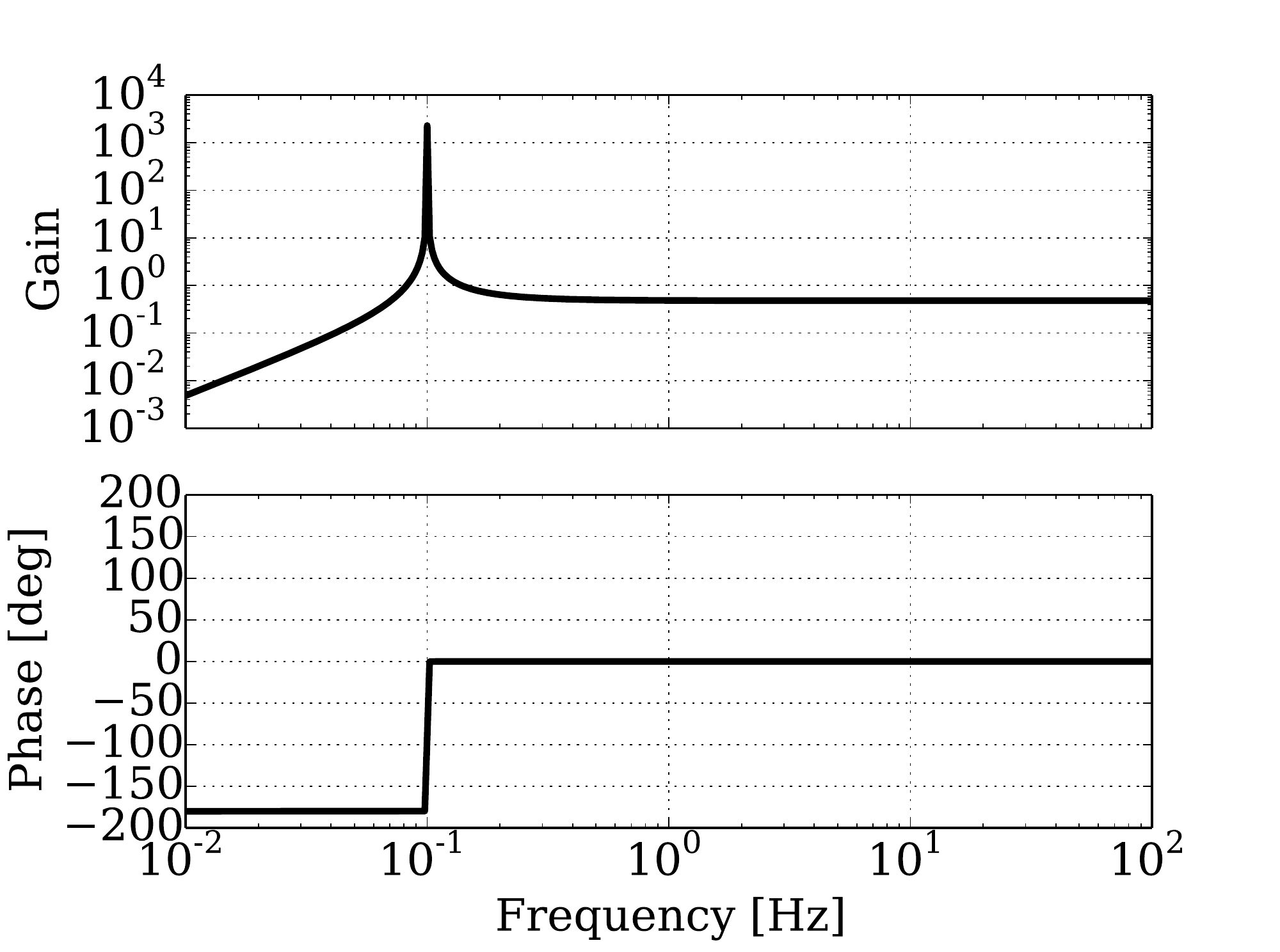}
\caption{\label{fig:calib}Calibration factor from the horizontal rotational angle to $s_1$, which is the GW amplitude equivalent signal derived from the horizontal rotation of the bars. Above the resonant frequency of the horizontal rotation at 0.1 Hz, the calibration factor is constant as derived from Eq. (\ref{eq:H}).}
\end{figure}

\begin{figure}[tb]
\centering
\includegraphics[width=70mm]{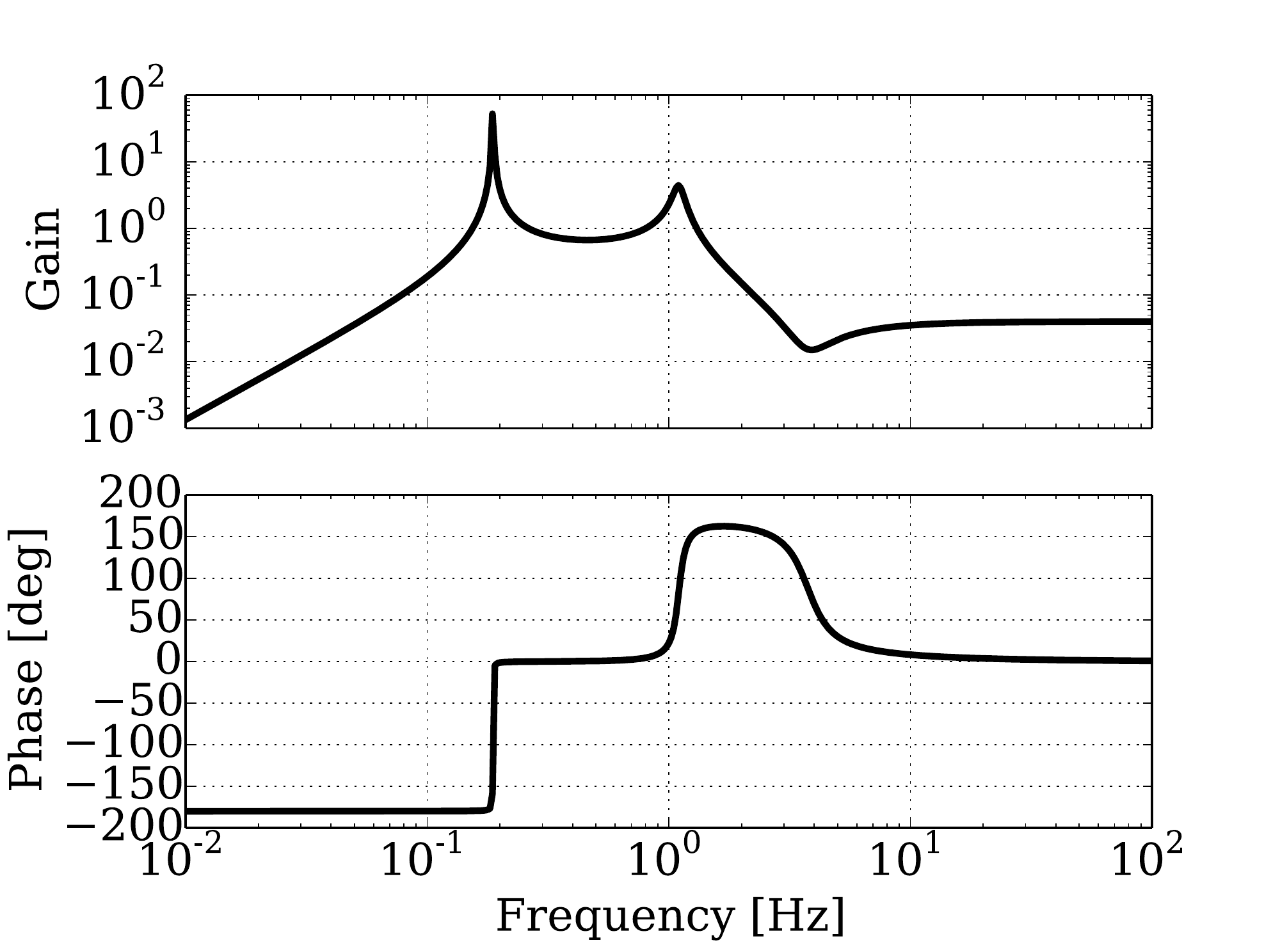}
\caption{\label{fig:calib2}Calibration factors from the vertical rotational angle to $s_{2,3}$, which is the GW amplitude equivalent signals derived from the vertical rotation of the each bar. The peak at 0.15 Hz is the resonant frequency of the vertical rotation of the bar. Above the the resonant frequency of the optical bench, which is the 1.1 Hz with the quality factor of 10, the calibration factor is decreased because the optical bench also rotates due to the GW.}
\end{figure}

In our case, the $H_{Ai}^{\rm TM}$ are
\begin{equation}
H_{\times i}^{\rm TM} \simeq \frac{q_{\times i}^{\rm TM}}{2I_i^{\rm TM}} = 0.48
\end{equation}
above their resonant frequencies, where $i=1,2$ and $3$.
For the optical bench,
\begin{equation}
H_{\times 1}^{\rm OB} \simeq 0,
\end{equation}
\begin{equation}
H_{\times 2,3}^{\rm OB} \simeq \frac{q_{\times 2,3}^{\rm OB}}{2I_{2,3}^{\rm TM}} = 0.44.
\end{equation}
$H_{\times 2,3}^{\rm OB}$ are not zero because the side view of the optical bench does not show four-fold symmetry, while $H_{\times 1}^{\rm OB}$ is derived to be zeros from $q_{\times 1}^{\rm OB} = 0$.
Therefore, the calibration factor for the $s_2$ and $s_3$ are about 10 times smaller than $s_1$ in our setup as shown in Figs. \ref{fig:calib} and \ref{fig:calib2} above 2 Hz.

\subsection{Sensitivity and noise sources}
\begin{figure}[tb]
\centering
\includegraphics[width=70mm]{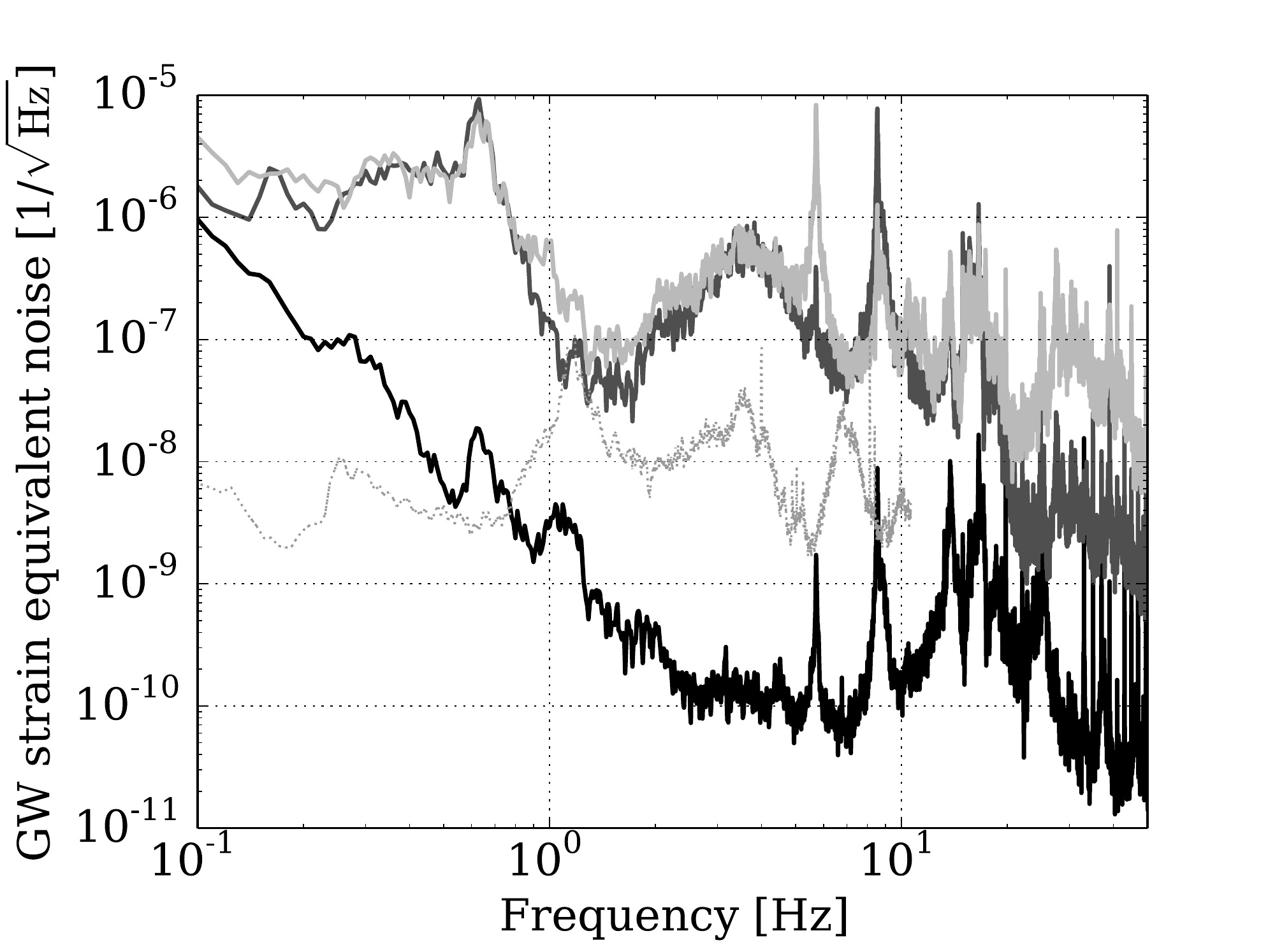}
\caption{\label{fig:strain}GW equivalent noise spectra obtained from the multi-output TOBA compared with the spectrum of the previous detector. The black, dark gray, and light gray lines are the spectra of $s_1, s_2$, and $s_3$, respectively. The dotted gray line represents the sensitivity of the previous detector \cite{Ishidoshiro2011}.}
\end{figure}
\begin{figure}[tb]
\centering
\includegraphics[width=85mm]{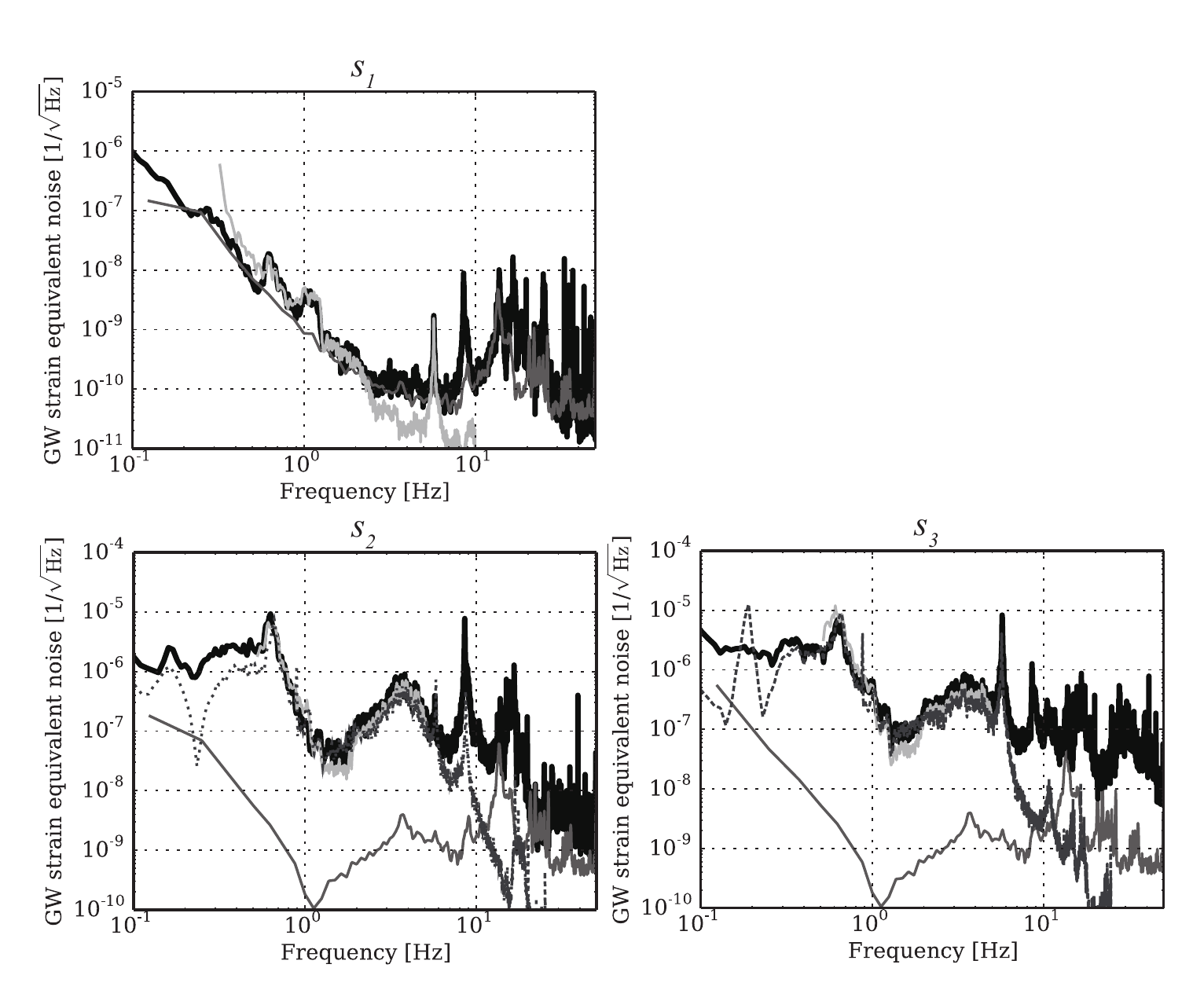}
\caption{\label{fig:sens_limits}GW equivalent noise spectra and the spectra of the respective noise sources. The left top, left bottom, and right bottom graphs are the sensitivities and the noise sources of $s_1$, $s_2$, and $s_3$, respectively.
The black lines are the GW equivalent noise spectra.
The dark gray lines show the readout noise measured with the test mass fixed to the optical bench.
The light gray blue lines are the seismic noise estimated from the transfer function from the ground to the sensor directly measured using the AVIT.
The dotted dark gray lines are the seismic noise calculated using the theoretical transfer function from the ground to the sensor.
The theoretically calculated seismic noise is not plotted in the left top, since the coupling mechanism of seismic noise to $s_1$ is unknown.}
\end{figure}

The solid lines in Fig. \ref{fig:strain} are the GW strain equivalent noise spectra obtained from the multi-output TOBA.
The dotted line is the sensitivity of the first prototype.
It shows that the sensitivity of $s_1$ that is represented in the black line in Fig. \ref{fig:strain} is improved by about 100 times at the maximum compared to the first prototype.
Figure \ref{fig:sens_limits} shows the dominant noise sources in the three signals.
The sensitivity of the Multi-output TOBA is limited mainly by the interferometer readout noise and the seismic coupling noise.
The readout noise, shown in the solid dark gray lines, is estimated from the readout signal measured with the test masses fixed on the optical bench.
It is considered to be induced from the fiber optics since the contribution from the other laser noise source, such as the intensity fluctuation and the frequency fluctuation, are lower than the measured readout noise.
The light gray lines are the seismic noise estimated from the motion of the AVIT table measured by the seismometers on the AVIT and the transfer functions from the seismometers to the laser sensors directly measured by exciting the AVIT.
While the coupling mechanism to $s_1$ is now under investigation, the seismic noises of $s_2$ and $s_3$ are induced because the translational seismic motions excite the vertical rotations of the test masses.
Since the heights of the suspension point and of the center of mass are different, the translational force on the suspension point applies the torque in the roll direction.
The dotted dark gray lines in the two bottom spectra in Fig. \ref{fig:sens_limits} are the seismic noise calculated from the theoretical transfer functions from the ground to the test masses.
The sensitivity curve, the seismic noise estimated from the measured transfer functions, and the theoretical seismic noise fit well in $s_2$ and $s_3$.
In addition to the large seismic coupling, the small calibration factor from the rotation to the GW amplitude, as derived in the previous section, worsen the sensitivity of $s_2$ and $s_3$ by about 1,000 times than $s_1$.

\begin{figure}[tb]
\centering
\includegraphics[width=60mm]{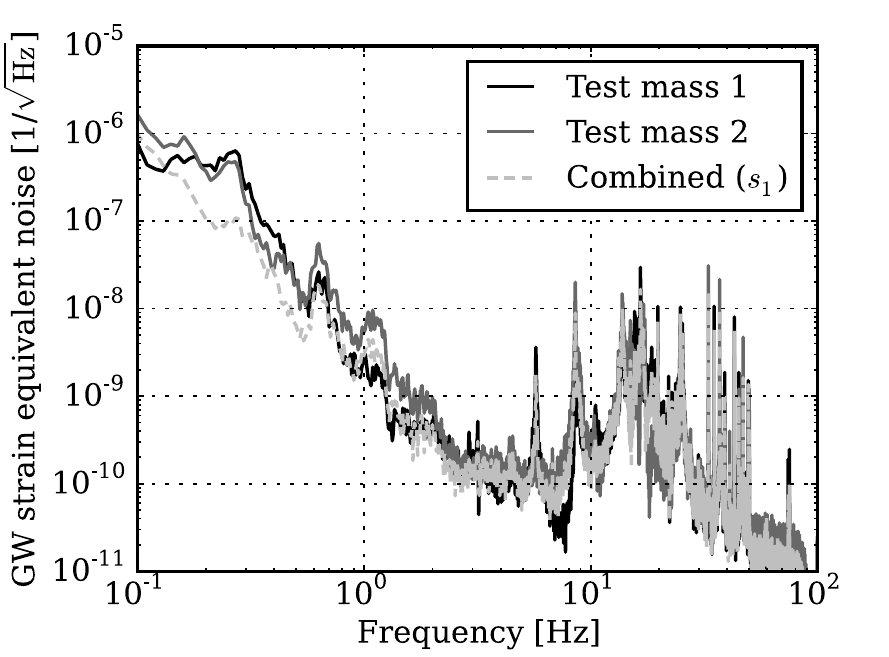}
\caption{\label{fig:CMR}Effect of the common mode noise rejection. The black and the dark gray lines are the strain sensitivities calculated using single test mass, 1 and 2, respectively. The dotted light gray line is the sensitivity derived by subtraction of the signals from the two test masses.}
\end{figure}

The performance of the common mode noise rejection between the two test masses is shown in Fig. \ref{fig:CMR}.
As described in the section \ref{sec:TM}, the centers of masses are designed to be at the same position since the noise may be reduced when the sensitivity is limited by the common rotational displacement.
However, subtraction of the two signals is not effective for the noise reduction in our case, since there are almost no coherence in two rotational signals.
The readout noise is not correlated between the two independent interferometers.
Also the seismic coupling noise is not correlated, since the seismic motion in $y$ direction couples to the rotation signal of the test mass 2, and the motion in $x$ direction couples to the signal of the test mass 1 in Fig. \ref{fig:TOBA}.
Since there are no correlation between the seismic motions in the orthogonal directions, the two rotational signals are not correlated.

\begin{figure}[tb]
\centering
\includegraphics[width=92mm]{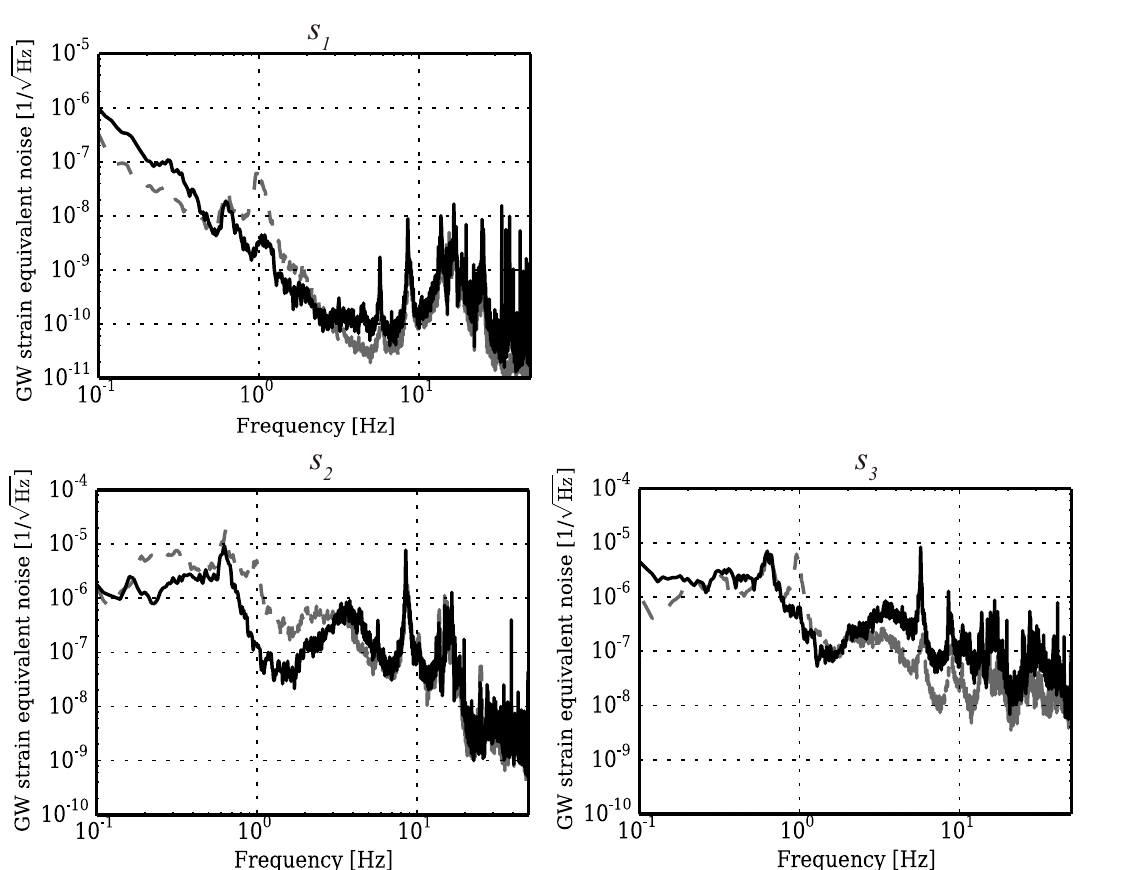}
\caption{\label{fig:AVITeffect}Sensitivities of signals with the AVIT on and off.
The top, left bottom and right bottom graphs represent the sensitivity of $s_1$; $s_2$, and $s_3$, respectively.
The black lines are the sensitivities with the AVIT on, while the dashed gray lines are the ones with AVIT off.}
\end{figure}
Figure \ref{fig:AVITeffect} shows the efficiency of the AVIT.
The dotted gray lines are the sensitivity measured without vibration attenuation using the AVIT, while solid black lines are the sensitivity measured with the AVIT working.
The sensitivity at around 1 Hz largely improved thanks to the AVIT.
At 4-10 Hz, the noise levels become worse by the AVIT.
It is supposed to be the control noise induced by the AVIT since the resonance of the support frame disturbs the AVIT control loops to have enough phase margin.

\subsection{Stability and Gaussianity}
We performed the observational run from 19:50 JST, December 10, 2014 to 19:50 JST, December 11, 2014. 
The observation system stably continued for more than 24 hours.

\begin{figure}[tb]
\centering
\includegraphics[width=92mm]{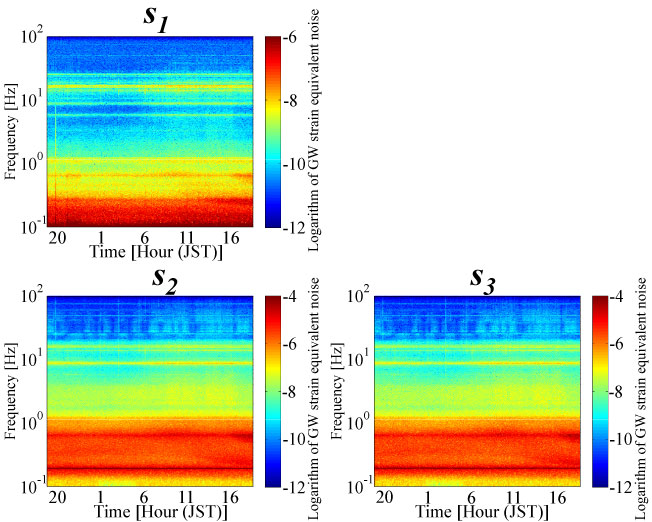}
\caption{\label{fig:spectrogram}(color online). Spectrograms of the power density of the three signals during the 24-hour observation from 0.1 to 100 Hz.}
\end{figure}

The spectrograms of the three signals during the observation are shown in Fig. \ref{fig:spectrogram}.
The noise levels are almost the same for 24 hours except for the earthquake which occurred at around 20:00, December 10.
\begin{figure}[tb]
\centering
\includegraphics[width=92mm]{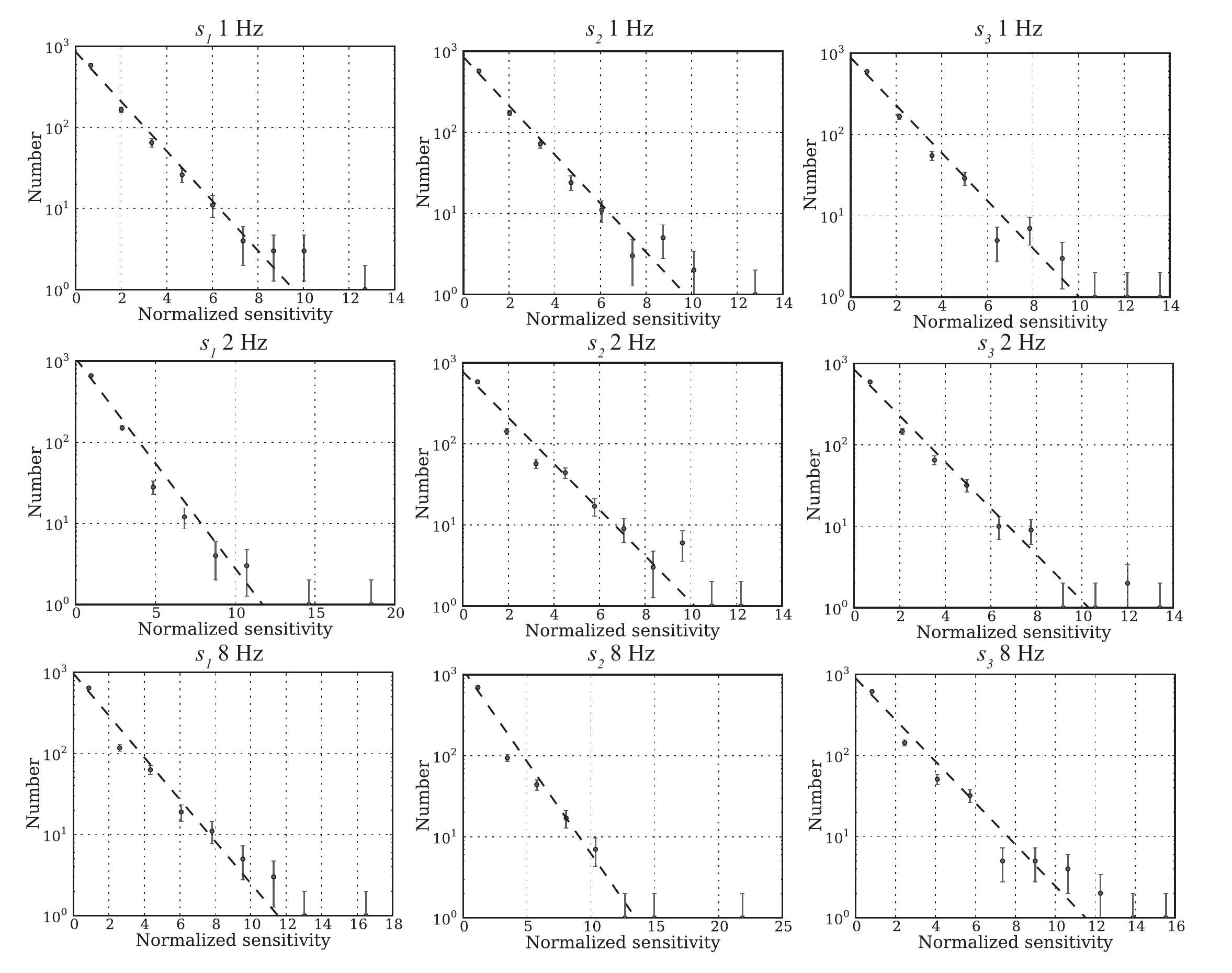}
\caption{\label{fig:gaussianity}Histograms of the normalized power spectral density of $s_1$, $s_2$, and $s_3$ at 1, 2, and 8 Hz. The dotted line at each panel corresponds to a Gaussian distribution.}
\end{figure}

The histograms of the power spectrum density of the obtained data are shown in Fig. \ref{fig:gaussianity}.
The columns represent the signals, $s_1$, $s_2$, and $s_3$, and the lines represent the frequencies of the collected data, 1 Hz, 2 Hz, and 8 Hz, respectively.
The sensitivities shown here are normalized by the average sensitivity over the whole observation.
The all three signals are distributed according to Gaussian distributions.

Using these data, we performed several GW search analysis: the search for continuous GWs, stochastic GW backgrounds, and intermediate mass black hole binaries.
Continuous GW signals had been searched for at 6-7 Hz and set an upper limit of $3.6 \times 10^{-12}$ on the dimensionless GW strain at 6.84 Hz \cite{Eda2016}.
Also, at 2.58 Hz, a new upper limit $\Omega_{\rm gw}h_0^2 < 6.0 \times 10^{18}$ is set on the energy of the stochastic GW background, where $\Omega_{\rm gw}$ is the GW energy density per logarithmic frequency interval in units of the closure density and $h_0$ is the Hubble constant in units of 100 km/s/Mpc \cite{Kuwahara}.
The GWs from the binary systems with the mass of 100 $M_{\odot}$, and 200 - 1000 $M_{\odot}$ were searched using the matched filtering method, and no signals had been discovered \cite{ShodaThesis, Sato2016}.
The continuous GW signals and GWs from the binary systems were searched at around 1-10 Hz for the first time. The upper limit on the energy of the stochastic GW background has also been updated using the detector with improved sensitivity.

\section{Discussion and Future Plan}
The sensitivity of TOBA was improved mainly due to the passive and active vibration isolation systems.
The pendulum suspension system is used as a passive vibration isolation in order to reduce the seismic motion above 1 Hz.
The vibration isolation system for the sensors, as well as for the test masses, is effective.
In addition, the AVIT are introduced in order to reduce the seismic motion around 1 Hz. 
While its performance was limited by the resonance of the frame where the AVIT is sitting and the actuation range of the PZTs, the seismic motion of the table is reduced by about 10 times from the ground in the translational three degrees of freedom.
The combination of those vibration isolation system improves the sensitivity at around 1 - 10 Hz by 10 - 100 times compared with the previous prototype TOBA.

Also, the three independent signals are successfully obtained simultaneously and stably as the multi-output signals, while the sensitivity of the signals from the vertical rotational motion are worse than that from the horizontal rotational motion. 
The sensitivity was different among the three signals because of the large coupling from the seismic motion and the cancellation of the GW signals between the test masses and the optical bench.

In order to improve the detector performance, careful optimization of the suspension design is necessary.
The reinforcement of the frame for the AVIT is required in order to gain more vibration attenuation ratio.
The shape of the optical bench should have four-fold symmetry in order to improve the sensitivities of $s_2$ and $s_3$ so that the optical bench would not react to GWs.
Also, the positions of the centers of masses of the test masses and the optical bench would be adjusted from the outside of the vacuum tank using moving masses and actuators, such as pico-motors, in order to search for the state that minimize the seismic coupling.
Such modification also helps to investigate the seismic coupling mechanism in the torsion pendulum further.
In addition to the suspension design, the readout system is required to be modified in order to improve the sensitivity.
Having more space on the optical bench would enable us to use the spacial laser beams instead of the fiber optics.

For further sensitivity improvement, it is necessary to upgrade several technologies.
Other than the seismic attenuation system developed in our paper, the low loss suspension system and the cryogenic system are critical path to achieve the final target sensitivity.
Also, the newly introduced observation method, the multi-output system, is required to be investigated further, though the demonstration was successful.
For example, reduction of the suspension thermal noise in the vertical rotational signals would be the one of the important subjects.
Also, it is necessary for the two vertical rotation signals to improve the detector configuration by the methods as explained in the previous section in order to achieve the same sensitivities as the horizontal rotation signal.
However, since the fundamental noise sources for the vertical rotation are the same as the ones for the horizontal rotation, the achievable sensitivities of $s_2$ and $s_3$ are expected to be almost same as $s_1$.

%%midterm
For the midterm upgrade, GW strain equivalent noise of $10^{-15}$ at 0.1 Hz, which is the sensitivity such that the gravity gradient effect, so called Newtonian noise, could be observed, would be the target.
Besides the further optimization for the seismic coupling reduction, the key technology for the mid-term upgrade would be the low-loss suspension for the reduction of the thermal noise in the midterm phase.
The gravity gradient effect is the noise caused by  the gravity perturbation due to the seismic motion, acoustic sound, motion of the object around the detector, and so on \cite{Saulson1984}.
It is necessary to establish the method of Newtonian noise canceling in this phase for the further sensitivity improvement below 1 Hz.
Conversely, precise detection of gravity gradient signals caused by earthquakes would also be applicable for the early alert of large earthquakes.
The gravity gradient signal from a large earthquake is considered to be used as the prompt alert of the earthquake since the gravity signal propagates at the speed of light, which is much faster than the seismic waves \cite{Harms2015}.
The full-tensor configuration realized by the multi-output system would also be effective in terms of the gravity gradient detection and cancellation \cite{harms2015newtonian}. 

%long term
For the long term upgrade to achieve the final target sensitivity, $1 \times 10^{-19}$ ${\rm Hz}^{-1/2}$, the 10-m scale large test mass is necessary.
The rotation of the test mass is required to be read by the Fabry-Perot interferometer in order to reduce the shot noise.
Also, the cryogenic system to reduce the thermal noise would be critical in order to improve the sensitivity at low frequencies.
With this sensitivity, intermediate mass black holes binaries at the luminosity distance of 10 Gpc would be observed with the signal to noise ratio of 5 \cite{Ando2010}.
Such observational results would be a key to reveal the evolution processes of stellar and supermassive black holes, globular clusters, and galaxies.

\section{Summary}
We have developed the new TOBA detector that employed the new suspension system for the multi-output system.
It successfully worked as the multi-output detector, i.e., the three independent signals was derived simultaneously from the single detector.
It demonstrates the new technique to improve the event rate and the parameter estimation accuracy.
Also, the sensitivity obtained from the horizontal rotational signal is improved from the previous prototype because of the passive and active vibration systems.
Especially, AVIT, which is the compact active vibration isolation system realized with the PZT actuators and seismometers, reduced the seismic displacement at lower frequencies than the resonant frequency of the pendulum.
AVIT is also expected to be effective for the vibration isolation for the heat link for the cryogenic system which is planned to be used in the future.
Those new technologies would be a new step towards the low-frequency GW astronomy.
While it is necessary to develop other technologies, such as a cryogenic system and a large system, observation of the binary black hole systems with multi-output TOBA would provide various astronomical information.

\section{Acknowledgment}
This work is supported by Grants-in-Aid from the Japan Society for the Promotion of Science (JSPS), JSPS Fellows Grants No. 26.8636 (K. E.), 24.7531, and 15J11064 (A. S.). This work also is supported by JSPS Grants-in-Aid for Scientific Research (KAKENHI) Grant No. 24244031 (M. A.), 24103005, 15H02082, and 15K05070 (Y. I.).

%%%%BibTex%%%%%%


\end{document}